\documentstyle[11pt,aaspp4]{article}
\begin{document}
\newcommand{\lya}{Lyman~$\alpha$}
\newcommand{\lyb}{Lyman~$\beta$}
\newcommand{\za}{$z_{\rm abs}$}
\newcommand{\ze}{$z_{\rm em}$}
\newcommand{\cmtwo}{cm$^{-2}$}
\newcommand{\nhi}{$N$(H$^0$)}
\newcommand{\nzn}{$N$(Zn$^+$)}
\newcommand{\ncr}{$N$(Cr$^+$)}
\newcommand{\degpoint}{\mbox{$^\circ\mskip-7.0mu.\,$}}
\newcommand{\halpha}{\mbox{H$\alpha$}}
\newcommand{\hbeta}{\mbox{H$\beta$}}
\newcommand{\hgamma}{\mbox{H$\gamma$}}
\newcommand{\kms}{\,km~s$^{-1}$}      
\newcommand{\minpoint}{\mbox{$'\mskip-4.7mu.\mskip0.8mu$}}
\newcommand{\mv}{\mbox{$m_{_V}$}}
\newcommand{\Mv}{\mbox{$M_{_V}$}}
\newcommand{\peryr}{\mbox{$\>\rm yr^{-1}$}}
\newcommand{\secpoint}{\mbox{$''\mskip-7.6mu.\,$}}
\newcommand{\sqdeg}{\mbox{${\rm deg}^2$}}
\newcommand{\squig}{\sim\!\!}
\newcommand{\subsun}{\mbox{$_{\twelvesy\odot}$}}
\newcommand{\et}{{\it et al.}~}

\def\ltsima{$\; \buildrel < \over \sim \;$}
\def\simlt{\lower.5ex\hbox{\ltsima}}
\def\gtsima{$\; \buildrel > \over \sim \;$}
\def\simgt{\lower.5ex\hbox{\gtsima}}
\def\arcs{$''~$}
\def\arcm{$'~$}
\title{SPECTROSCOPY OF LYMAN BREAK GALAXIES IN THE HUBBLE DEEP FIELD\altaffilmark{1}}
\author{\sc Charles C. Steidel\altaffilmark{2,3}}
\affil{Palomar Observatory, California Institute of Technology, Mail
Stop 105-24, Pasadena, CA 91125}
\affil{e-mail: ccs@astro.caltech.edu}
\author{\sc Mauro Giavalisco\altaffilmark{4}}
\affil{Observatories of the Carnegie Institution of Washington, 813 Santa
Barbara Street, Pasadena, CA 91101}
\affil{e-mail: mauro@ociw.edu}
\author{\sc Mark Dickinson}
\affil{Space Telescope Science Institute, 3700 San Martin Drive, Baltimore, MD
21218}
\affil{e-mail: med@stsci.edu}
\author{\sc Kurt L. Adelberger}
\affil{Palomar Observatory, California Institute of Technology, 
Mail Stop 105-24, Pasadena, CA 91125}
\affil{e-mail: kla@astro.caltech.edu }

\altaffiltext{1}{Based on obervations obtained at the
W.M. Keck Observatory, which is operated jointly by the  
California Institute of Technology and the University of California, and
with the NASA/ESA Hubble Space Telescope, which is operated by AURA, Inc., under
contract with NASA}
\altaffiltext{2}{Alfred P. Sloan Foundation Fellow}
\altaffiltext{3}{NSF Young Investigator}
\altaffiltext{4}{Hubble Fellow}

\begin{abstract}
We report on the initial results of a spectroscopic investigation of 
galaxies in the Hubble Deep Field which 
exhibit spectral
discontinuities between the F450W and F300W passbands, indicative
of the presence of the Lyman continuum break in the redshift
range $2.4 \simlt z \simlt 3.4$.  We have employed
color selection criteria similar to those we have used for
selecting high redshift galaxy candidates from ground--based
images. We find that, as for the ground--based color selection,
the criteria are very successful in selecting high redshift
objects. Of the 8 galaxies observed (selected from a list
of 23 candidates with magnitudes equivalent to ${\cal R}\le 25.3$), 
5 have confirmed redshifts in the
range $2.59 \le z \le 3.22$, with the remaining 3 being
indeterminate because of contamination from nearby brighter objects. 
As expected, the HST filter system
is sensitive to a somewhat broader range of redshifts than
our ground--based $U_n G {\cal R}$ filter system,  and
therefore the surveyed volume per unit area on the sky is correspondingly larger.
The distribution of candidates
on the plane of the sky is clearly non--uniform, consistent with
the available ground--based data on the high redshift galaxies.
Most Lyman break objects in the Hubble Deep Field exhibit a similar
range of morphological properties to the $z>3$ galaxies we
have previously identified in other fields, characterized by
very compact cores (some with multiple components)
with half--light radii of $0.2-0.3$ arc seconds, 
often surrounded by more diffuse and asymmetric ``halos''. 
A few of the brighter HDF Lyman break galaxies, however,
have particularly unusual morphologies.   

\end{abstract}

\section{ INTRODUCTION}

The ``Hubble Deep Field'' (hereinafter ``HDF'') (Williams \et 1996) presents an
opportunity to assess the colors and morphologies of galaxies down to
unprecedentedly faint magnitude levels. 
Given that
the field was observed across a wide color baseline, including
the UV F300W filter, it is a natural place to extend our ongoing studies
of the galaxy populations at $z\sim 3$ (Steidel \et 1995, 1996; Giavalisco
\et 1996; hereinafter Papers III, IV, V) that have been based on the Lyman 
continuum break entering 
the UV passband at substantial redshifts, giving the objects
colors which distinguish them from the rest of the faint field galaxies . 
Our ground--based
photometric system and selection criteria (see Steidel \et 1993, 1995 for details) 
have been shown to be sensitive to galaxies in the redshift range
$3.0 \le z \le 3.5$, based on direct spectroscopic follow--up of the
$z>3$ candidates using the W. M. Keck telescope (Paper IV). It is possible to use
very similar selection criteria-- essentially ``flat'' rest--UV spectra across
the observed--frame optical passbands, with a dramatic ``drop--out'' in the
ultraviolet passband-- in the HST/HDF photometric system to pick out the
candidate high redshift galaxies in the Hubble Deep Field. In this paper,
we report the first spectroscopic observations of these ``Lyman break'' galaxies 
in the HDF. We have also performed a morphological analysis of these bright
candidates following the lines of our previous work presented in Paper V. 

\section {SAMPLE SELECTION}

Given the very small amount of time (a matter of only a few days) between the
availablility of the Hubble Deep Field data and our observing run on the 
W. M. Keck telescope, galaxies were selected as Lyman break candidates
in a manner that was not necessarily optimized for the HST filter system.
We simply adopted criteria very similar to those used in our ground--based
survey. All of the magnitudes on the HST
system were converted to ``AB'' magnitudes (such that a galaxy with
equal magnitudes in each passband has a spectrum that is ``flat'' ,
i.e. $f_{\nu} \propto \nu^0$). We then approximated our ground--based
${\cal R}$ passband, which has an effective wavelength of 6930 \AA\ for
a flat--spectrum source, by averaging the F606W and F814W AB magnitudes. 
For the colors of the objects of interest, this is likely to be
very close to a ${\cal R}$ magnitude in both normalization and effective
wavelength. (We will call this magnitude ${\cal R}$ hereinafter). We then formed
the colors $F300W - F450W$ and $F450W - {\cal R}$, and applied
a ``spectral curvature'' criterion, 
$$F300W - F450W > 1.2 + (F450W-{\cal R}),
$$ 
which is equivalent to requiring that the break across the two bluer passbands
is more than 3 times the break across the two redder ones. In addition, we
expect that the unabsorbed continua of high redshift galaxies will be very
blue (even with the blanketing effect of the Lyman $\alpha$ forest, up to
redshifts of $z \sim 3.5$), so that we require that $F450W -{\cal R} < 1.2$. 
In practice,
essentially all of the objects which satisfied the ``spectral curvature''
criterion also satisfied the ``blueness'' criterion. All of the colors were
measured using the isophotal aperture defined in the Version 1 catalog produced
by the HDF team; such apertures are conservatively large for objects as bright
as those considered here, and are certainly not optimal for obtaining colors
with the highest S/N. 

We then took the entire
catalog of objects having ${\cal R} \le 25.3$ (this has been shown
to be the practical limit for follow--up spectroscopy with Keck, and also
ensures that the limits in F300W for Lyman break objects will be extremely
robust) and applied the color selection criteria. Figure 1 shows the
two--color diagram for all of the HDF objects with ${\cal R} \le 25.3$, together
with the region of this plane satisfying the above color criteria.
The filtering resulted
in a sample of 25 resolved objects culled from the three detectors of the
WFPC2 camera, and one stellar object. Two of the resolved objects were rejected 
as spurious after close inspection of the HDF images. The colors of the 
stellar object may have been slightly
affected by saturation; a long slit spectrum obtained on the same night
as the slit mask observations described below show it to be a subdwarf
star. All of the candidates for follow--up spectroscopy, with accompanying
photometry and positions, are listed in Table 1\footnote{In the original
generation of the list of candidates for spectroscopy, an error was made
in the conversion of the F606W magnitude to the AB scale, so that
the $F450W - {\cal R}$ color was measured to be 0.2 magnitudes bluer, and
the ${\cal R}$ magnitude 0.2 magnitudes fainter, than the values that
are now tabulated. In addition, when the catalog F300W magnitudes did not
exceed 1 $\sigma$ above sky in the isophotal aperture, we assigned lower
limits on the F300W magnitude equivalent to $+1\sigma$ above sky--in some
cases this made the $F300W - F450W$ break less pronounced than in the
original list. As a result, there are objects in Table 1 which no longer
satisfy the adopted color criteria completely. We have retained all of
the original candidates for spectroscopy in Table 1, however.}.  
A mosaic of all 23 of the galaxies is shown in Figure 2. 

We expected {\it a priori} that the color selection applied to the HST
data set would be sensitive to somewhat lower redshifts than our
ground--based criteria, since the Lyman limit of a galaxy is well
within the F300W passband for any redshift $z \simgt 2$. To better quantify
the expected redshift range encompassed by our selection criteria, 
we convolved the throughput curves of the
HST passbands with the same model galaxy template (Bruzual and Charlot 1993;
see Paper III) we have used in our ground--based system, accounting for
blanketing in the Lyman alpha forest following the prescriptions of Madau (1995)
and by assuming that photons shortward of the Lyman limit in the galaxy rest frame
are completely absorbed.   This assumption is probably valid, since recent
{\it Hopkins Unltraviolet Telescope} observations of nearby star
forming galaxies have shown no detectable flux emerging 
shortward of the Lyman limit [Leitherer \et 1995], 
and because of the known high opacity of
intergalactic H~I for redshifts $z \simgt 2.5$ (cf. Paper III, Madau 1995). 
The model galaxy colors, which should be considered rough estimates,
satisfy our criteria for candidate selection in the redshift range
$2.4 \le z \le 3.4$, i.e. a much broader range of redshifts than in our
ground--based high redshift galaxy selection ($3.0 \le z \le 3.5$). 
This then explains (in part) the
surprisingly large number of Lyman break candidates; whereas we would
have expected $\sim 2$ Lyman break galaxies in the 4.7 square arc minute
HDF field
of view to ${\cal R}=25$ based on our ground--based statistic of
$0.40$ galaxies per arc square minute (see Paper IV) 
we see 10 such candidates in the HDF. After accounting for the probed volume,
which is larger by a factor of $\sim 2$ than the equivalent volume probed
by the $U_nG{\cal R}$ selection criteria (because of the larger redshift range),
and the fact that one probes
a few tenths of a magnitude fainter in the star--forming galaxy luminosity
function at the average redshift probed, these numbers are probably consistent
with one another within the errors.  A detailed analysis of the luminosity function 
of Lyman break galaxies in the HDF is postponed to a separate paper.

\section {OBSERVATIONS}

The spectroscopic observations were obtained on the night of 22 January 1996 (UT) 
on the W. M. Keck telescope with the Low Resolution Imaging Spectrograph (Oke \et 
1995).  Two slit masks were constructed, each containing slits for 8 of the
Lyman break candidates in the HDF, plus 15--20 additional slits for
galaxies in the HDF ``flanking fields'', which were much shorter
(1--2 orbit) HST exposures. The candidate objects for spectroscopy were assigned
slits based on a weighting scheme that was largely subjective, but was loosely
based on apparent magnitude (brighter objects being given larger weight), and
efficiency of how many slits could be placed on candidates within the HDF as
a function of position angle, etc. The first mask, which is the one actually used
for observations, was optimized to have a slit falling roughly along the major axis
of one particularly interesting galaxy, C4-06, which also happens to be the brightest
candidate in the HDF, and to allow another slit to be placed on C4-09, another
interesting candidate with a peculiar morphology.    

The weather was far from optimal on the night the observations were made, with
variable cirrus and mediocre seeing ($\sim 1.3-1.5$ arc seconds FWHM). We 
obtained a total of 7200s of integration through the first slit mask, in
six separate exposures of 1200s each. (This integration represents $\sim 90$\% of
the workable time we experienced in the entire two night observing run.) 
The telescope was moved small amounts ($\sim 2$\arcs)
parallel to the slit direction  between each exposure, so that the spectrum of each
object fell at 3 independent spatial
positions on the Tektronix 2048 $\times$ 2048 detector. We note that this integration
time is only about half that obtained for the $z > 3$ galaxy spectra which were 
presented in Paper IV. The spectra were obtained with a 300 line/mm 
grating blazed at 5000 \AA\ in first order, resulting in a spectral 
resolution of $\sim 12$ \AA\ through the 1\secpoint 4 slits. 

Two of the slits assigned to HDF Lyman break candidates (C4-02 and C2-01 in Table 1)
were contaminated by light from nearby mask alignment star holes, so that no useful
data were obtained for them. A third galaxy, C3-01, is contaminated by
two nearby galaxies at smaller redshifts (see Figure 2) due to the poor seeing and
the unfavorable alignment of the slit on the plane of the sky. We have obtained
redshifts for the remaining 5 candidate $z>2.4$ galaxies, and indeed they all
fall within the expected range of redshifts.  

The spectra of the five confirmed high redshift galaxies in the HDF
are presented in Figure 3.

\section{THE SPECTRA}

The spectra generally span the same general range of properties as those 
we have obtained previously in other fields at similar redshifts-- generally (but not
always) weak or absent Lyman $\alpha$ emission, sometimes very strong
interstellar absorption lines, and when the S/N permits, stellar
absorption features of Si~IV, C~IV, and He~II (cf. Paper IV). 
Again, the present spectra
are generally of inferior quality (as compared to those described in Paper IV)
due to the adverse weather conditions experienced
during the 1996 January Keck observing run.   

We draw attention in particular to the
spectrum of HDF C4-06, which is a high surface--brightness, but highly
elongated galaxy. The interstellar absorption features in this galaxy,
at $z=2.803$, are stronger than for any other high redshift galaxy
we have obtained thus far; they are comparable in strength to 
the lines in the spectrum of the extremely bright ``protogalaxy'' candidate
found serendipitously by Yee \et (1996). It is clearly an usual
object morphologically (see \S 5), as most of the color--selected $z \sim 3$ galaxies
are quite compact and relatively ``round'' (Paper V).
C4-06 has an aspect ratio of about 5:1, and given the 
elongated morphology, the strength of the interstellar lines 
lends credence to a dynamical interpretation of the line widths, 
as we have suggested in Paper IV. The strongest interstellar lines have
equivalent widths of $\sim 5$\AA\ in the rest frame, implying a velocity field of
more than 1000\kms FWHM. Unfortunately, although the object was observed
with the slit oriented along the major axis of the galaxy, the seeing
was too poor to determine whether this velocity field has a substantial
shear or rotation component. 

HDF C4-09, at $z=3.226$,  is another object with a somewhat unusual morphology. It is
evidently comprised of 4 distinct, compact components
of similar luminosity and separated by only $\sim 1$\arcs\ on the plane
of the sky. Aside from the {\it number} of individual compact components
comprising this object, it is not unlike several of the $z>3$ galaxies
presented in Paper V which exhibit these ``multiple compact'' morphologies. 
(see also \S V).  We note the apparently very strong Si~IV feature in
the spectrum in Figure 3, as well as the moderately strong
($W_{\lambda}=20$ \AA) Lyman $\alpha$ emission line. 
There is some evidence for a velocity difference in the Lyman $\alpha$
line in the two-dimensional spectrum of this object, suggesting that
it is probably not the gravitationally lensed images of a single object.
There is also no obvious sign of a lensing object to very faint apparent
magnitude levels. 

HDF C4-08, at $z=2.591$, has the strongest Lyman $\alpha$ emission (of any 
non--AGN)
of the high $z$ objects we have observed spectroscopically to date, with
a rest--frame Lyman $\alpha$ equivalent width of $\sim 60$\AA. Morphologically,
it is very similar to objects which exhibit weak or absent Lyman $\alpha$ (see
\S V and Figure 2).

HDF C2-05, at $z=2.845$, has very weak Lyman $\alpha$ emission superposed 
on a strong absorption that is more evident in the two--dimensional
sky--subtracted spectrum than in the 1--d spectrum shown in Figure 3. The
redshift is supported by several of the low--ionization interstellar lines
that are found to be a common feature of the high redshift galaxy spectra
in the far--UV.  

HDF C3-02, at $z=2.775$, is the least secure of the 5 redshifts presented,
and is also the faintest galaxy in terms of apparent magnitude. We base
the redshifts mainly on features shortward of 5000 \AA, where there are
few strong features in the sky, and where apparent absorption features
are most likely to be real. However, there are relatively strong
apparent absorption features which agree with some of the expected stellar
lines (e.g., C~IV, Si~IV, He~II) at longer wavelengths. 
We plan to obtain additional data on this
galaxy in the future to make the redshift assignment more secure.

\section{THE MORPHOLOGY OF THE $z>2.4$ GALAXIES}  

We have performed an analysis of the light profiles for 
the HDF Lyman break galaxies 
as we have done previously for our $z > 3$ $U_n G {\cal R}$ selected
galaxies observed with {\it HST} (Paper V).
Figure 4 shows radial
profile plots of each of the galaxies. Table 2 contains a summary
of the measured parameters of the light profiles for each object.

Broadly speaking, and not surprisingly, the HDF Lyman break galaxies
are overall very similar in their morphological properties to our
previous sample; they are generally characterized by compact morphologies     
which we have described as consisting of a high surface--brightness
``core'', surrounded by much lower surface brightness nebulosities
or ``halos''.  
The cores, which often have a central surface brightness of $\sim 22-23$ 
mag arcsec$^2$, are typically resolved and have sizes of $\sim 0.5$--$1$ 
arcsec, corresponding to $\sim 3.6$--$7.3h_{50}^{-1}$ (6.1--12.2) kpc with 
$q_0=0.5$ (0.05), with half-light radii of $\sim 0.2$--$0.3$ arcsec, or 
$\sim 1.5$--$2.2h_{50}^{-1}$ (2.4--3.7) kpc. The ``halos'' are larger in size 
and are generally irregularly distributed around the cores. The ``cores'', 
which have the roughly the 
same sizes as the present-day bulges of spiral galaxies and cores of 
ellipticals, contain a very large fraction of the UV light (and therefore
of the star formation) of the galaxies. As we have argued in Paper V,
the compact morphologies do not simply result from the well--know
surface brightness selection effects inherent in the {\it HST} images, 
since our sample of galaxies selected using ground--based data
is overall very similar morphologically. 

It is interesting to note that the observed central surface
brightnesses are very high; the typical 22-23 mag. arcsec$^{-2}$ corresponds
to a rest--frame surface brightness in the $B$ band of 17.7-18.7 mag. 
arcsec$^{-2}$ if it is assumed that the spectrum remains essentially
flat to rest--frame $B$ (this assumption is supported by 
infrared photometry of a subsample of the $z>3$ galaxies presented
in Paper IV) and that the typical redshift is $z\sim 2.8$.  
These numbers are at least ten times higher than
the central surface brightness of a typical spiral disk at the
present epoch (Freeman 1970).

As in Paper V for the ground--based  
$U_nG{\cal R}$ selected galaxies, we have quantitatively examined the 
morphology of the HDF galaxies in two ways. To analyze their average radial 
light profile we fit elliptical isophotes to the {\it HST} images. One should 
fit the center, ellipticity and position angle of each individual isophote. 
However, the presence of the substructures and the relatively small number 
of pixels that comprise the image make this method inaccurate, mostly due 
to the fluctuations in the centers of each isophote. We adopted a procedure 
wherein we used the center derived from the inner isophote 
(typically $r<0.25''$), and keeping this center fixed, we fit ellipticities 
and position angles for each isophote. This technique differs slightly
from that adopted in Paper V, where we kept
ellipticity and position angle fixed to an average value, because the HDF
images are considerably deeper and have better angular resolution (thanks to
the sub-pixel dithering and processing) than our previous {\it HST} images. 
In practice, we fit each azimuthally--averaged radial profile with
both an $r^{1/4}$ and an exponential law to determine which of the two best
describes each galaxy. 
Again, the fits are intended to broadly classify the light profiles and are
not intended to imply that a particular galaxy rigourously follows a given model.
In several cases it is clear that neither analytic profile provides a very
good fit, and in such cases the tabulated $r_e$ parameters are clearly
without significance. 
The half--light radii, as in Paper V, were determined from growth curves
and not from the parameters of the fitted profiles.

Notes on individual objects:

C2-03: Although this galaxy appears to be elongated, it is only the small
knot at the center of the circle in Figure 2 that has the colors expected
for a galaxy at $z > 2.4$. This object would almost certainly have been
missed in a ground--based search because of the confusion with the
foreground object. 

C2-05 and C2-06: Although we have a spectrum of only one of this fairly
close pair (the separation is only about 2\secpoint5), they have very
similar broadband colors (see Table 1), making it likely that they
are at similar redshifts. 

C2-07: Here again, it is only the single compact object at the center of the
circle in Figure 2 which has the appropriate color signature to be a Lyman
break object. This is another case that might have been missed at ground--based
resolution.  

C4-02, C4-03: A second, fainter compact region is observed in these
galaxies. 

C4-04: Again, only the object at the center of the circle in Figure 2 has 
the expected colors of a galaxy with $z > 2.4$.

C4-05: Another example of a ``double core'' object.

C4-06: As noted above in \S 4, this is the brightest among the
Lyman break candidates in the HDF, and also the most morphologically
peculiar. It is the only high redshift object found using Lyman break color
selection that we have seen, in either the HDF or in the fields
we studied previously in Paper V, which exhibits a significantly
elongated morphology. As can be seen from Table 2, the two ``blobs''
comprising this object are of very similar apparent magnitude and
central surface brightness. It is not beyond the realm of possibility
that this object is gravitationally lensed (it is very close to what appears to be a
relatively high redshift early--type galaxy). 

C4-08: Although this object has much stronger Lyman $\alpha$ emission
than any of the Lyman break galaxies we have studied spectroscopically
to date, it is not appreciably different from typical in terms
of its light profile. 

C4-09: This object bears some resemblance to an ``Einstein cross'' gravitational
lens configuration. However we consider it more likely that it is just
an example of a ``multiple compact core'' morphology such as we have
seen in several instances both in the HDF and in the fields presented
in Paper V. There is no evidence for a foreground galaxy centered
on the 4 compact components. As discussed in \S 4, there is also
some evidence that the components have slightly different redshifts.
These are probably 4 sub--units in the process of merging, or 4 bright
star-forming regions within a single, underlying galaxy.  

\section{DISCUSSION}

We have obtained confirming spectra of 5 out of 23 galaxies from
the 3 Wide Field Camera chips of the Hubble Deep Field, selected to have
${\cal R} \le 25.3$ and to satisfy color criteria essentially identical
to those we have used successfully in our ground--based survey for
very high redshift galaxies. The
{\it HST} data are, as expected, sensitive to a larger range of redshifts
than the ground--based $U_nG{\cal R}$ photometric system; the 5 redshifts
all fall within the expected range $2.4 \le z \le 3.4$ based on the simple
model used (see Madau \et 1996 for a color prescription optimized for the
{\it HST} filter system). Thus, the Lyman break technique for isolating
high redshift galaxies is again shown to be extremely efficient.

It is perhaps significant that in the sample of 5 redshifts available thus far,
3 are within $\pm 2600$ \kms of one another, near redshift $z \approx 2.80$. 
We also point out  
the distribution of the candidates among the three WFC chips, with
12 on WF2, 9 on WF4, and only 2 on WF3. These two findings indicate
the possible presence of large--scale clustering in the distribution of
actively star--forming galaxies at high redshifts. 
Such inhomogeneities and
signals of very large scale clustering are also present in our larger--field
ground--based data as well, as we will discuss in future work. 

We reiterate a conclusion which was reached in Paper V-- that the high redshift
galaxies are in general very compact, with scales comparable to the cores
of present--day luminous galaxies. The bulk of the star formation at high $z$
is occuring in very compact regions, of very high surface brightness, and
which generally exhibit a relatively high degree of azimuthal symmetry. 

We view the spectra presented in this paper as further evidence of the
effectiveness of using the Lyman continuum break criterion for isolating
well-defined populations (essentially volume limited) of high redshift
galaxies. There is a great deal of science that may be done, from studies
of the luminosity function of galaxies at extreme redshifts, to studies
of large--scale structure, using the high resolution images and extremely
accurate colors afforded by the Hubble Deep Field data alone. These avenues
are all being explored at the moment.  

\acknowledgements

We would like to thank the W.M. Keck Foundation for making the telescope a reality,
and Bev Oke and Judy Cohen 
for providing the instrument, that together make it possible
to obtain spectra of heretofore hopelessly faint galaxies. We also thank Bob Williams
for investing his discretionary time into the Hubble Deep Field, which is sure
to yield important scientific results for years to come, and we thank
the entire HDF team who worked so hard to bring the data to the public
in an immediately useful form. We are grateful to Max Pettini for
his comments on a draft of the paper. C.S. acknowledges support
from the National Science Foundation grant AST-9457446, and from the Alfred
P. Sloan foundation. M.G. acknowledges support from the Hubble Fellowship
program through grant number HF-0107.01-94A, awarded by the Space Telescope
Science Institute, which is operated by the Associated Universities for
research in Astronomy, Inc., under NASA contract NAS5-26555.

\newpage
\begin{deluxetable}{lcccccc}
\tablewidth{0pc}
\scriptsize
\tablecaption{Lyman Break Candidates in the HDF, ${\cal R} <25.3$}
\tablehead{
\colhead{Object} & \colhead{${\cal R}$} & \colhead{F450W$-{\cal R}$} & 
\colhead{F300W$-$F450W} &
\colhead{$\alpha$(J2000)} &
\colhead{$\delta$(J2000)} & \colhead{z}  } 
\startdata
C2-01 &   25.19 & 0.95 & $>$2.25 &  12  36  45.30 &  62  13  47.89 &   \nodata \nl
C2-02 &   24.30 & 0.13 & 1.87 &  12  36  44.01 &  62  14  10.81 &   \nodata \nl
C2-03 &   25.00 & 0.33 & 1.77 &  12  36  45.84 &  62  14  12.73 &   \nodata \nl
C2-04 &   25.21 & 0.33 & 1.67 &  12  36  50.22 &  62  13  30.74 &   \nodata \nl
C2-05 &   23.58 & 0.41 & 1.74 &  12  36  48.23 &  62  14  17.59 &   2.845 \nl
C2-06 &   24.35 & 0.33 & 1.65 &  12  36  48.16 &  62  14  19.42 &   \nodata \nl
C2-07 &   25.17 & 0.15 & 1.96 &  12  36  49.93 &  62  13  52.99 &   \nodata \nl
C2-08 &   24.84 & 0.38 & 2.37 &  12  36  50.02 &  62  14  02.08 &   \nodata \nl
C2-09 &   25.18 & 0.76 & $>$2.44 &  12  36  51.13 &  62  13  49.77 &   \nodata \nl
C2-11 &   25.21 & 1.07 & $>$2.13 &  12  36  52.66 &  62  13  40.06 &   \nodata \nl
C2-12 &   24.77 & 1.05 & $>$2.15 &  12  36  53.34 &  62  13  30.39 &   \nodata \nl
C2-13 &   24.60 & 0.21 & 1.89 &  12  36  54.98 &  62  13  48.05 &   \nodata \nl
C2-14 &   21.08 & 0.98 & 2.79 &  12  36  52.73 &  62  14  33.04 & star  \nl
C3-01 &   24.37 & 0.33 & 1.63 &  12  36  54.63 &  62  13  15.82 &   \nodata \nl
C3-02 &   24.89 & 0.10 & 1.62 &  12  36  51.24 &  62  12  28.34 &   2.775 \nl
C4-01 &   25.30 & 0.30 & 2.54 &  12  36  49.59 &  62  12  20.76 &   \nodata \nl
C4-02 &   25.12 & 1.11 & $>$2.09 &  12  36  48.54 &  62  12  16.86 &   \nodata \nl
C4-03 &   25.25 & 0.25 & $>$2.95 &  12  36  46.84 &  62  12  27.13 &   \nodata \nl
C4-04 &   25.23 & 0.87 & $>$2.33 &  12  36  48.21 &  62  11  46.94 &   \nodata \nl
C4-05 &   24.89 & 0.26 & $>$2.94 &  12  36  43.18 &  62  12  39.92 &   \nodata \nl
C4-06 &   23.48 & 0.86 & $>$2.39 &  12  36  45.26 &  62  11  53.78 &   2.803 \nl
C4-07 &   25.22 & 0.37 & 1.72 &  12  36  42.32 &  62  12  33.56 &   \nodata \nl
C4-08 &   24.37 & 0.51 & $>$2.57 &  12  36  41.58 &  62  12  39.92 &   2.591 \nl
C4-09 &   24.21 & 1.11 & $>$2.24 &  12  36  41.14 &  62  12  04.00 &   3.226 \nl
\enddata
\end{deluxetable}
\newpage
\begin{deluxetable}{lccccccccc}
\tablewidth{0pc}
\scriptsize
\tablecaption{Morphological Parameters}
\tablehead{
\colhead{Name} & 
\colhead{Redshift} &
\colhead{$m_{0.2}$\tablenotemark{a}} &
\colhead{$m_{0.7}$\tablenotemark{b}} &
\colhead{$m_{i}$\tablenotemark{c}} &
\colhead{Fit\tablenotemark{d}} &
\colhead{$r_{e,0}$\tablenotemark{e}} & 
\colhead{$r^C_{1/2}$\tablenotemark{f}} &
\colhead{$r^T_{1/2}$\tablenotemark{g}} & 
\colhead{a/b} }
\startdata
C2-01  &  \nodata  & 26.23 & 25.30 & 25.19 & e & 0.19 & 0.24 & 0.25 & 1.53 \nl
C2-02  &  \nodata  & 25.33 & 24.79 & 24.30 & d & 1.13 & 0.22 & 0.27 & 1.46 \nl
C2-03  &  \nodata  & 25.99 & 25.11 & 25.00 & e & 0.56 & 0.24 & 0.25 & 2.13 \nl
C2-04  &  \nodata  & 25.82 & 25.21 & 25.21 & e & 0.17 & 0.16 & 0.16 & 2.12 \nl
C2-05  & 2.845 & 24.23 & 23.61 & 23.58 & d & 0.80 & 0.16 & 0.16 & 1.23 \nl
C2-06  &  \nodata  & 25.46 & 24.35 & 24.35 & d & 0.38 & 0.27 & 0.27 & 1.19 \nl
C2-07  &  \nodata  & 25.51 & 25.17 & 25.17 & p?& \nodata  & 0.13 & 0.13 & 2.00 \nl
C2-08  &  \nodata  & 25.69 & 24.84 & 24.84 & d & 0.25 & 0.22 & 0.22 & 1.64 \nl
C2-09  &  \nodata  & 26.12 & 25.27 & 25.18 & d & 0.72 & 0.22 & 0.25 & 1.81 \nl
C2-11  &  \nodata  & 25.49 & 25.13 & 25.21 & p?& \nodata  & 0.13 & 0.13 & 1.16 \nl
C2-12  &  \nodata  & 25.57 & 24.77 & 24.77 & d & 0.29 & 0.20 & 0.20 & 1.86 \nl
C2-13  &  \nodata  & 25.25 & 24.60 & 24.60 & e & 0.32 & 0.18 & 0.18 & 1.20 \nl
C3-01  &  \nodata  & 24.74 & 24.37 & 24.37 & p?& \nodata  & 0.13 & 0.13 & 1.56 \nl
C3-02  & 2.775 & 25.83 & 24.99 & 24.89 & d & 1.73 & 0.25 & 0.25 & 2.13 \nl
C4-01  &  \nodata  & 25.79 & 25.30 & 25.30 & e & 0.10 & 0.16 & 0.16 & 1.42 \nl
C4-02  &  \nodata  & 25.65 & 25.15 & 25.12 & e & 0.08 & 0.16 & 0.16 & 1.20 \nl
C4-03  &  \nodata  & 25.93 & 25.25 & 25.25 & e & 0.13 & 0.18 & 0.18 & 1.91 \nl
C4-04  &  \nodata  & 26.05 & 25.23 & 25.23 & d & 0.98 & 0.22 & 0.22 & 2.12 \nl
C4-05  &  \nodata  & 25.71 & 24.99 & 24.89 & e & 0.19 & 0.22 & 0.22 & 1.29 \nl
C4-06a & 2.803 & 24.99 & 24.13 & 23.48 & e & 0.17 & 0.22 & 0.55 & $\sim 2.3$ \nl
C4-06b & 2.803 & 25.41 & 24.25 & 23.48 & e & 0.54 & 0.29 & 0.65 & $\sim 2.3$ \nl
C4-07  &  \nodata  & 25.56 & 25.22 & 25.22 & p?& \nodata  & 0.13 & 0.13 & 1.41 \nl
C4-08  & 2.591 & 25.42 & 24.70 & 24.37 & d & 0.66 & 0.20 & 0.27 & 2.91 \nl
C4-09  & 3.226 & 25.33 & 24.28 & 24.21 & d & 0.15 & 0.36 & 0.38 & 1.45 \nl
\enddata
\tablenotetext{a}{Magnitude in a $0.2$ arcsec radius aperture, on the $AB$ scale}
\tablenotetext{b}{Magnitude in a $0.7$ arcsec radius aperture, on the $AB$ scale}
\tablenotetext{c}{Isophotal magnitude, on the $AB$ scale}
\tablenotetext{d}{Function which better models the radial profile: d=$r^{1/4}$;
e=exp; p=point-source}
\tablenotetext{e}{Fit scale length: $r_e$ for $r^{1/4}$; $r_0$ for exp. In arcsec}
\tablenotetext{f}{Core half-light radius in arcsec, from direct photometry}
\tablenotetext{g}{Total half-light radius in arcsec, from direct photometry}
\end{deluxetable}
\newpage

\begin{figure}
\figurenum{1}
\plotone{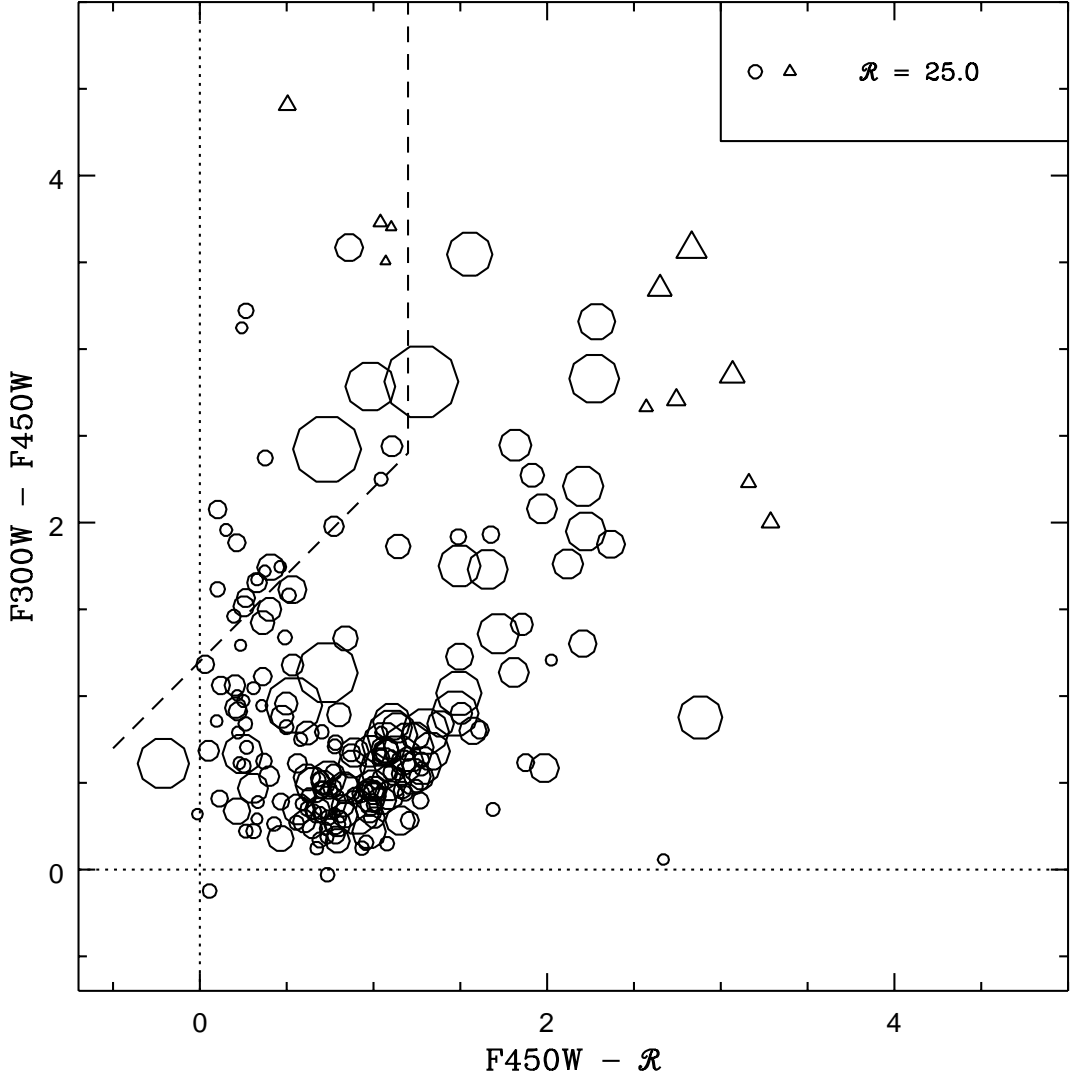}
\caption{
A two--color diagram showing all galaxies in the HDF with ${\cal R} \le 25.3$; the
region within which we considered a galaxy to be a candidate in the redshift
range $2.4 \le z \le 3.4$ is indicated, to the left and above the dashed lines.
The circles are objects detected in all three bands, and the triangles are
objects with only lower limits on the F300W magnitude. The sizes of both symbols 
scale inversely with ${\cal R}$ magnitude.  }
\end{figure}
\newpage
\begin{figure}
\figurenum{2}
\plotone{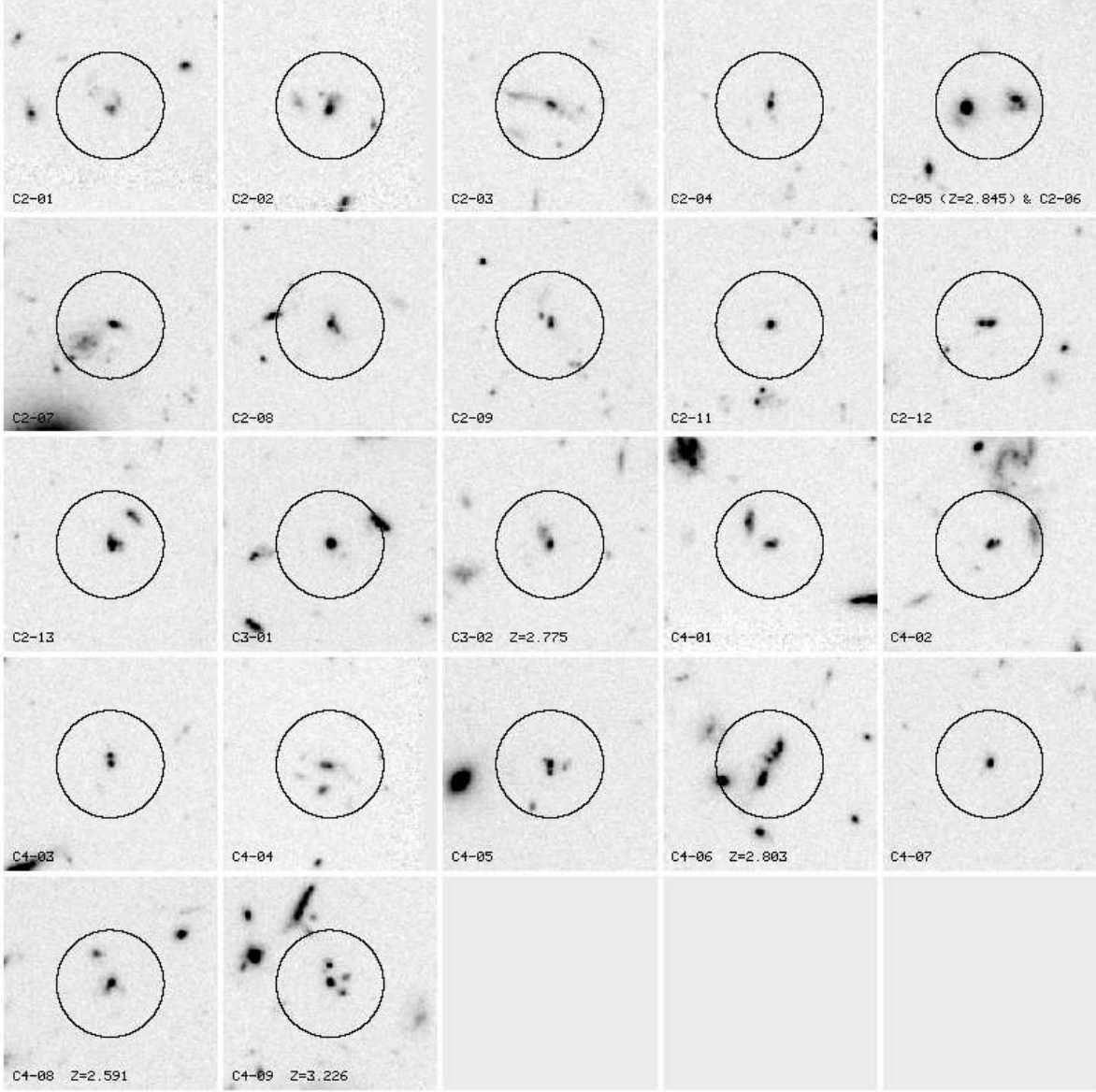}
\caption{Mosaic showing the candidate $2.4 \le z \le 3.4$ Lyman break
galaxies in the Hubble Deep Field, selected to have ${\cal R} \le 25.3$.
The images shown are the weighted sum of the F606W and F814W  ``drizzled''
HDF frames. Each box is 10\arcs\ on a side. The objects of interest are at the
centers of the circles (see \S 5 for notes on individual objects).}
\end{figure}
\begin{figure}
\figurenum{3}
\plotone{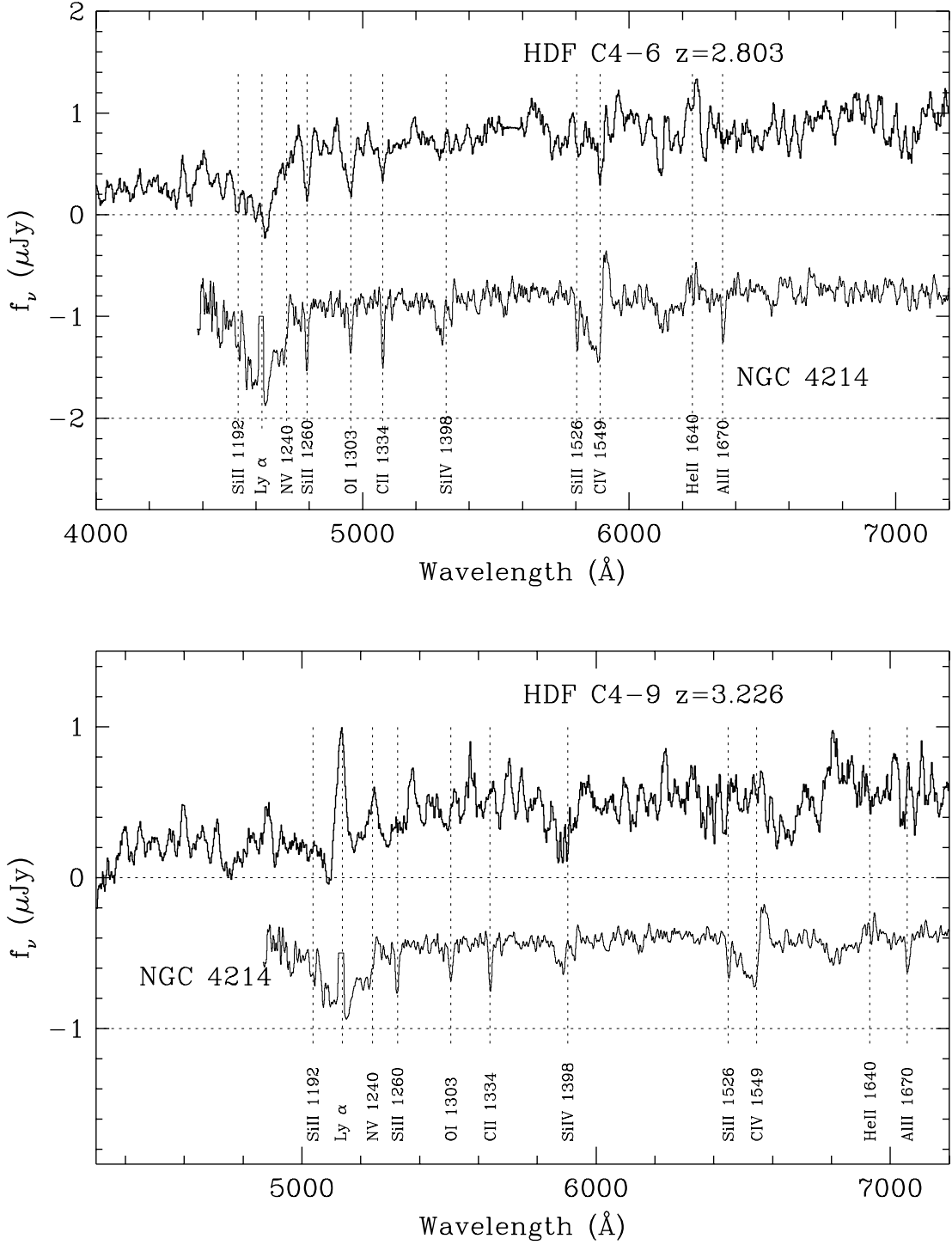}
\caption{Spectra of five of the Lyman break galaxies in the Hubble Deep
Field.  The identified
redshifts are indicated. Plotted below each spectrum is a scaled spectrum
of the nearby starburst galaxy NGC 4214 (Leitherer \et 1996), with the wavelength 
scale shifted to match that of the high redshift galaxy. The positions of a number
of stellar and interstellar features which are commonly observed in
both nearby and distant star-forming galaxies are marked. 
}
\end{figure}
\begin{figure}
\figurenum{3 (cont.)}
\plotone{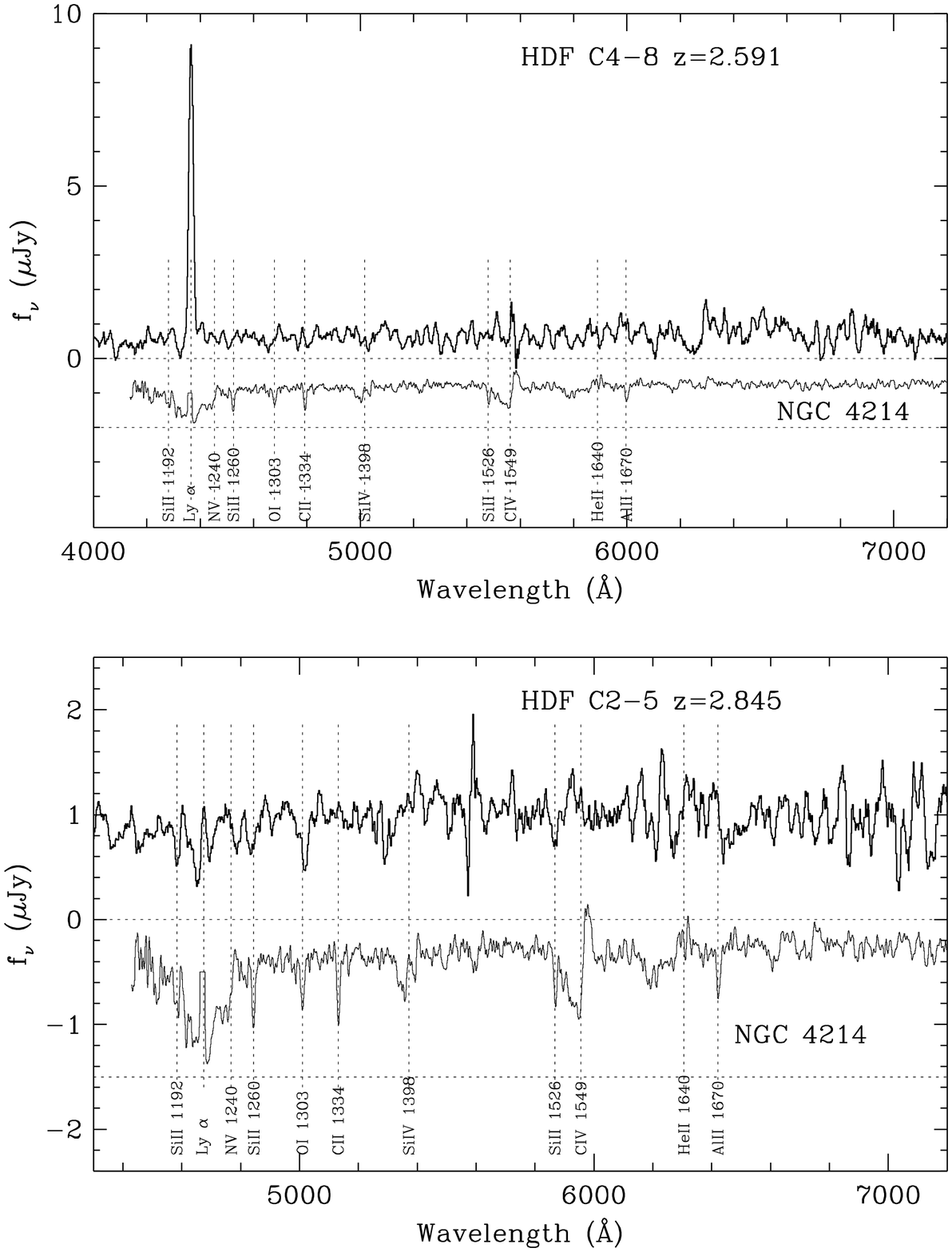}
\caption{}
\end{figure}
\begin{figure}
\figurenum{3 (cont.)}
\plotone{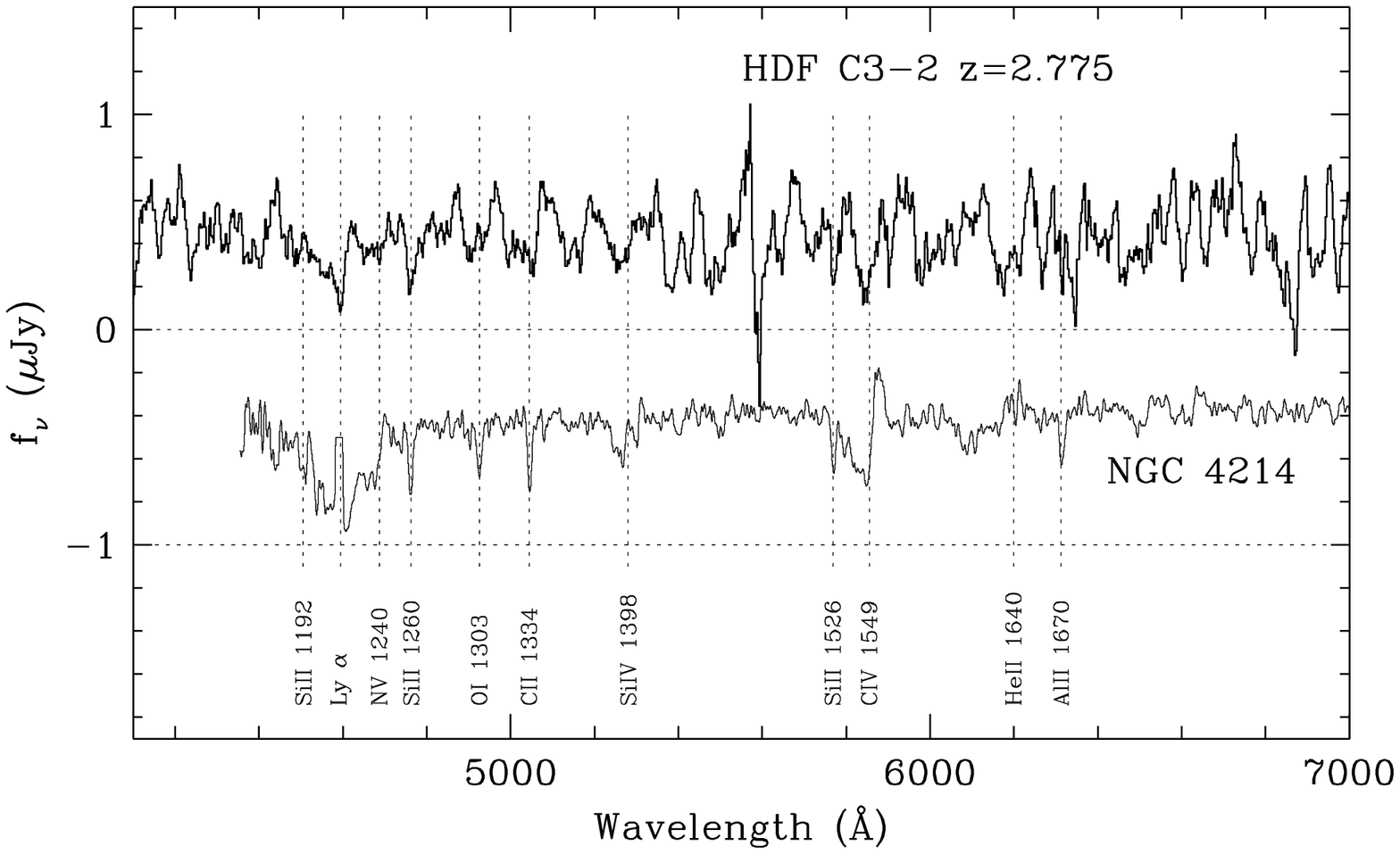}
\caption{}
\end{figure}
\begin{figure}
\figurenum{4}
\plotone{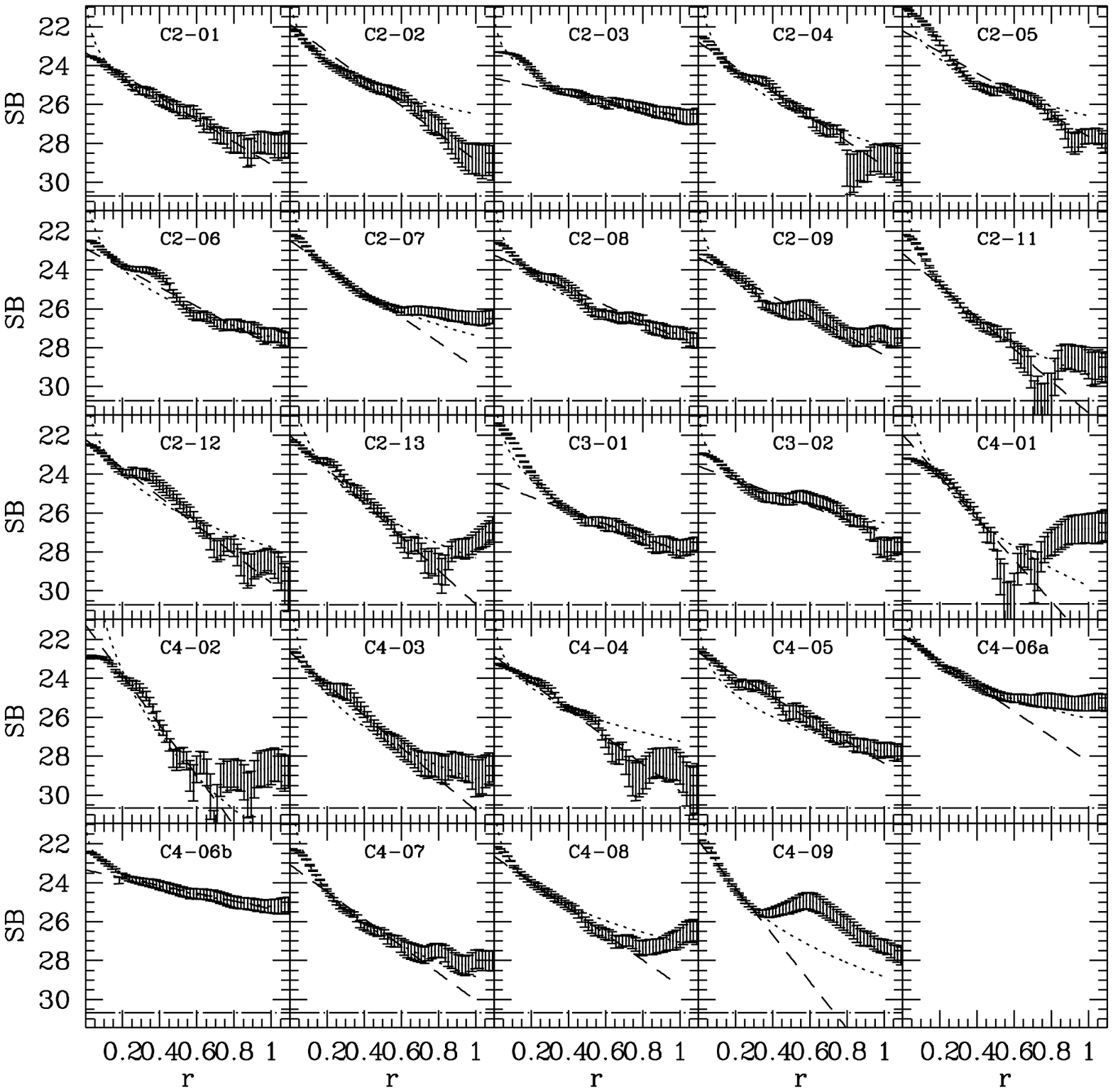}
\caption{Radial profiles of the galaxies identified as having $2.4 \le z \le 3.4$
and ${\cal R} \le 25.3$. The long-dash curves are $r^{1/4}$ fits, and the
dotted curves are exponentials. The surface brightness is in AB magnitudes at an 
effective observed--frame wavelength of 7000\AA, per
square arc second. }      
\end{figure}

\end{document}